\documentclass[twocolumn,showpacs,preprintnumbers,amsmath,amssymb]{revtex4}
\usepackage{graphicx}
\usepackage{dcolumn}% Align table columns on decimal point
\usepackage{bm}% bold math
\usepackage{amssymb}
\usepackage{amsmath}
\usepackage{amsthm}
\usepackage{latexsym}

\newcommand{\be}{\begin{equation}}
\newcommand{\ee}{\end{equation}}
\newcommand{\bea}{\begin{eqnarray}}
\newcommand{\nn}{\nonumber}
\newcommand{\eea}{\end{eqnarray}}
\newcommand{\bn}{\boldsymbol{\nabla}}

\begin{document}
\title{Tracing the geometry around a massive, axisymmetric body to measure, \\
through gravitational waves, its mass moments and electromagnetic moments.}

\author{Thomas P.~Sotiriou}
 \email{tsotiri@phys.uoa.gr}

\author{Theocharis A.~Apostolatos}%
 \email{tapostol@cc.uoa.gr}
\affiliation{Section of Astrophysics, Astronomy, and Mechanics\\
Department of Physics \\
National and Kapodistrian University of Athens\\
Panepistimiopolis, Zografos GR-15783, Athens, Greece.
}%
\begin{abstract}
The geometry around a rotating massive body, which carries charge and electrical currents could be described by
its multipole moments (mass moments, mass-current moments, electric moments, and magnetic moments).
When a small body is orbiting around such a massive  body, it will move
on geodesics, at least for a time interval that is short with respect to the characteristic time of the binary
due to gravitational radiation. 
By monitoring the waves emitted by the small body we are actually tracing the geometry
of the central object, and hence, in principle,
we can infer all its multipole moments. This paper is a generalization of previous similar results by
Ryan. In his paper Ryan explored the mass and mass-current moments of a stationary,
axially symmetric, and reflection symmetric, with respect to its equatorial plane, metric,
by analyzing the gravitational waves
emitted from a test body which is orbiting around the central body in nearly circular equatorial orbits.
In our study we suppose that the gravitating source is endowed with intense electromagnetic field as well.
Due to its axisymmetry the source is characterized now by four families of scalar multipole moments:
its mass moments $M_l$, its mass-current moments $S_l$, its electrical moments $E_l$ and its magnetic moments
$H_l$, where $l=0,1,2,\ldots$. Four measurable quantities, the energy emitted by gravitational
waves per logarithmic interval of frequency, the precession of the periastron, the precession of
the orbital plane, and the number of cycles emitted per logarithmic interval of frequency,
are presented as power series of the newtonian orbital velocity of the test body.
The power series coefficients are simple polynomials of the various moments. If any of these
quantities are measured with sufficiently high accuracy, the lowest  moments, including
the electromagnetic ones, could be
inferred and thus we could get valuable information about the internal structure of the compact massive
body. The fact that the electromagnetic moments of spacetime can be measured demonstrates that
one can obtain information about the electromagnetic field purely from gravitational wave analysis.
Additionally, these measurements could be used as a test of the no-hair theorem for black holes.
\end{abstract}

\pacs{04.25.Nx, 04.30.Db}

\maketitle

%%%%%%%%%%%%%%%%%%%%%%
%%%%%%%%%%%%%%%%%%%%%%
\section{\label{sec:1}Introduction}
%%%%%%%%%%%%%%%%%%%%%%
While, it is well known for quite some time that the geometry of the vacuum around a massive object
is related directly to the  multipole moments \cite{Geroch,Hansen,Simon1,Simon2} of the central object, as in newtonian gravity,
Ryan \cite{Ryan} was the first who attempted to map the spacetime geometry of a central axisymmetric
body, through its mass and mass-current moments, on a few measurable physical quantities that are related to
the kinematics of a hypothetical test body that is orbiting around the central object while emitting
gravitational waves. Later, Ryan \cite{Ryan2} used the outcome of his work
to perform  the analysis on the output of gravitational wave detectors, in order to extract
the moments of the central body around which a much lower-mass body is orbiting and emitting waves.

Fortunately, we live at an era where technology may give us the opportunity
to observe astrophysical phenomena that are related to the highly distorted geometry of a compact
massive central object. Especially, the detection of gravitational waves from binaries will provide us with
data that are strongly dependent on the  geometry itself. Apart from the Earth-based detectors
which are restricted to detect moderate mass binaries (e.g., 1 to 300 $M_\odot$ for LIGO) \cite{LIGOetal},
the Laser Interferometer Space Antenna, LISA, when it will fly in space and start operating,
is expected to explore the geometry of very massive objects with exceptional high precision \cite{LISA}.
Besides, telescopes with higher and higher resolution, operating at various regions of the electromagnetic spectrum,
are able to probe the close neighborhood  of compact
massive objects, as in accretion disks and jets of AGN's.
All this information which is related, by one way or another,
to  the motion of small objects in the curved geometry of massive astrophysical objects could
somehow  shed light on the internal structure of the central object. Of course, it is not expected
to fully determine its structure from the knowledge of its moments  (this is not possible even in
Newtonian gravity). However, knowing the moments of the central body,
could set restrictions on the various models that are assumed to describe the
interior of the central body. Moreover, the no-hair theorem in the case of a large black hole at the role
of the central object could be fully tested, if
from our moment-extraction formulae we get the values of electric and magnetic moments.

The binaries we have considered in our paper are idealized with regard to the following
quite realistic assumptions:

(i) The central object is assumed to be stationary and axisymmetric, and
characterized by reflection symmetry with respect to its equatorial plane.
This is expected to be true for a quiescent massive object with
internal fluid motions that are strictly toroidal. The axisymmetry gives as the freedom
to describe the spacetime geometry with scalar, instead of tensorial, multipole moments
(see Ref.~\cite{Hansen}). The same is true in Newtonian gravity as well, although in that case
all other moments except the mass moments do not show up in the expansion of the gravitational potential.
Since we are taking into account the electromagnetic content of the central object
as well, we are considering four families of multipole moments to
characterize the geometry around the central body: Its mass moments $M_0, M_2, M_4,\ldots$,
its mass-current moments $S_1,S_3,\ldots$, its electric charge moments $E_0,E_1,E_2,\ldots$, and
its magnetic moments $H_0,H_1,H_2,\ldots$. Especially, the $M_0\equiv M$ moment is the mass of the object,
$S_1$ is its angular momentum, $E_0\equiv E$ is its charge, and $H_{1}$ is its magnetic dipole moment. In every family of moments, each moment
appears in steps of 2, and this holds good for the electromagnetic moments as well \cite{explain}.
This property is due to the reflection symmetry of
the metric itself (cf., \cite{Hansen}). The geometry of such objects can be described by the
Papapetrou metric which consists of only two dynamical functions. The third one (see Section \ref{sec:2})
could be easily inferred from the first two.

(ii) Although we plan to extend our exploration in a generic geodesic motion around such a
central object, in the present paper we only take into consideration nearly equatorial and nearly circular
geodesic orbits of test bodies in the fixed geometry of a central massive object. We know that
gravitational radiation from a test body that is far from its innermost stable circular orbit
tends to circularize the orbit \cite{Math,ApoKenPoi}, and therefore the orbit could safely be considered
circular if it has sufficiently long time to evolve
without being perturbed by other objects. Also, we know that at least for not extremely fast
rotating Kerr black holes the evolution of non-equatorial orbits due to radiation reaction is
such that their inclination remains almost constant while their radius decreases \cite{Hugh}.

(iii) The energy emitted in the form of gravitational radiation
will be assumed to be given by the quadrupole formula,
since there is no known way to fully analyze the wave emission in a generic geometrical background.
Furthermore, we assume that this energy is carried away by waves at infinity, and there is no
energy loss through any horizon, or due to thermal heating of the surface of the central object from the impact of
gravitational waves.

The rest of the paper is organized as follows: First, in Sec. \ref{sec:2} we define the observable quantities
that will be used to measure the moments of the metric.
These quantities are the periastron precession  $\Omega_\rho$, the precession of the orbital plane $\Omega_z$,
the energy emitted at infinity per logarithmic frequency change $\Delta E/\mu$, and the number of cycles
of the primary gravitational waves per logarithmic frequency change $\Delta N$. Especially the latter one,
which can be measured with high accuracy by the broad-band wave detectors that are operating now,
or will be built in the near future, is computed assuming that the phase of the waves is coming simply
from the dominant frequency, $f=\Omega/\pi$.
We also show how one defines the
mass and electromagnetic multipole moments of an axially symmetric body with reflectional symmetry, and
how these moments uniquely determine the metric of the space around the object. For the equatorial
plane, and slightly out of this plane, we write the metric as a power series of the Weyl radial coordinate $\rho$.
The coefficients of the power series are polynomials of the various moments of spacetime.
We end up this section by explaining the implications of reflection symmetry of the metric on the
electromagnetic fields.
Having in hand all these expressions that connect the observable quantities to the metric and hence to
the moments, we proceed in Sec. \ref{sec:3} to write down expressions for
the four astrophysical quantities, as power series of $v \equiv (M \Omega)^{1/3}$,
where $M$ is the mass of the central object and $\Omega$ is the orbital frequency of the test
body, observed at infinity.
This quantity is the newtonian orbital velocity of the orbiting
body, and is a measure of the gravitational field strength.
Following the analysis of Ryan \cite{Ryan}, we present
the power expansion of  $\Delta N$ with coefficients that include only the leading order contribution
of each central-body multipole moment. It is argued that these are the only terms we could get without
getting into the complicate analysis of wave emission. Finally, in Sec.~\ref{sec:4} we comment on
how the gravitational wave analysis could inform us about the moments of the central object, the accuracy with which
these moments could be computed, and the implications they could have on the observational verification
(or not) of the full no-hair theorem (when charges are included).
Throughout the paper units are chosen so that $G=c=1$.

%%%%%%%%%%%%%%%%%%%%%%
%%%%%%%%%%%%%%%%%%%%%%
\section{\label{sec:2}Observable quantities and moments}
%%%%%%%%%
%%%%%%%%%%%%%%%%%%%%%%

In this section we briefly present and discuss the measurable quantities
that Ryan has used as tracers to measure moments. A thorough presentation
and analysis of them can be found in Ref.~\cite{Ryan}.
Then, we present all formulae that determine the various moments and relate  them
with the metric describing the geometry around a central object.
Finally, we discuss what kind of electromagnetic fields are consistent with
the symmetries assumed for the metric.

%%%%%%%%%%%%%%%%%%
\subsection{\label{sec:2.1}
Quantities that can be measured through
 gravitational-wave analysis}
%%%%%%%%%%%%%%%%%%

As is explained in \cite{Ryan} there are four physical quantities in a binary with high-mass-ratio,
that can, in principle, be measured through the gravitational radiation emitted by the binary,
and are straightforwardly related to the spacetime metric of the massive body. It is exactly these
quantities that we will use here
as a basic information in order to extract the various multipole moments of the central object.
These are: (i) $\Omega_\rho$, the periastron precession of the low-mass body, (ii) $\Omega_z$,
the orbital-plane precession of the low-mass body, (iii) $\Delta E$, the energy emitted as
gravitational waves per logarithmic
interval of frequency, and most important (iv) $\Delta N$, the number of gravitational wave cycles
per logarithmic interval of frequency. The first two quantities are computed by analyzing
the geodesic motion of nearly circular, nearly equatorial orbits of a test body on a fixed
background metric, and can be measured
through the modulation they induce on the gravitational waves emitted by the binary.
The third one, although it is directly related to gravitational radiation emitted by the
binary, it can be easily inferred by the functional relation of the energy of the
test body, which changes adiabatically, to its orbital frequency,
which  is simply half the primary wave frequency.
Finally, the fourth one is the best measurable quantity, since the phase matching used in
data analysis leads to a highly accurate estimation of the frequency dependence of phase,
assuming detection has been established. The computation of this quantity
involves a number of approximations, since it is directly related to the  mechanism of gravitational
wave emission on a complicated metric background.

Here, for the sake of completeness, we rewrite the expressions of Ryan \cite{Ryan}, that relate
all these observable quantities with the metric without further comments on how these are computed.
Both precession frequencies are given by
\bea
\label{omegas}
\Omega_\alpha &=& \Omega-
\left( -\frac{g^{\alpha\alpha}}{2} \left[
(g_{tt} + g_{t \phi} \Omega)^2
    \left(\frac{g_{\phi \phi}}{\rho^2} \right)_{,\alpha \alpha} \right. \right. {} \nn \\
&- &  2 (g_{tt} + g_{t \phi} \Omega)(g_{t \phi} + g_{\phi \phi} \Omega)
    \left(\frac{g_{t \phi}}{\rho^2} \right)_{,\alpha \alpha} \nn \\
&+ & \left. \left. (g_{t \phi} + g_{\phi \phi} \Omega)^2
    \left(\frac{g_{t t}}{\rho^2} \right)_{,\alpha \alpha} {} \right] \right) ^{1/2} ,
\eea
where $\alpha$ stands for $\rho$, or $z$. Actually, the frequencies written above correspond to the
difference between the orbital frequency and the frequency of perturbations in $\rho$, or $z$, since
these differences are expected to show up in gravitational waves as a modulating frequency. Of course these
frequencies are accurate only for orbits that are slightly non circular and slightly non equatorial,
otherwise the frequencies would depend not only on the metric but on specific characteristics
of the orbit, its eccentricity and inclination.

The energy per unit test-body mass for an equatorial circular orbit in an axially symmetric spacetime is
\be
\label{Eovermu}
\frac{E}{\mu}=\frac{ -g_{t t} - g_{t \phi} \Omega}
{\sqrt{-g_{t t}-2 g_{t \phi}\Omega - g_{\phi \phi} \Omega^2} } ,
\ee
and thus, the specific energy released as gravitational radiation per logarithmic interval of frequency
is
\be
\label{delE}
\frac{\Delta E}{\mu}=-\Omega \frac{d (E/\mu)}{d \Omega}.
\ee
The expression above assumes that all the energy lost from the test body has been emitted at infinity
as radiation, and is neither wasted as thermal energy on the fluid of the central object, nor is it
``lost'' through any horizon.

The number of gravitational-wave cycles spent in a  logarithmic interval of frequency is
\be
\label{delN}
\Delta N=\frac{f \Delta E(f)}{dE_{\textrm{wave}}/dt},
\ee
where $dE_{\textrm{wave}}/dt$ is the gravitational-wave luminosity, which is assumed to
be exactly the rate of energy loss of the orbiting test body. As Ryan has analytically shown
the main contribution of $dE_{\textrm{wave}}/dt$ comes from the mass quadrupole radiative moment of
the binary. More specifically, up to fourth order of $v \equiv (M\Omega)^{1/3}$ after the leading order,
the gravitational wave luminosity is accurately computed from the quadrupole formula
\be
\label{dEdtI}
\left. \frac{dE_{\textrm{wave}}}{dt} \right|_{I_{ij}}=\frac{32}{5} \mu^2 \rho^4 \Omega^6
\ee
plus a contribution of the current quadrupole radiative moment, due to the motion of the central object around the
center of mass,
\bea
\label{dEdtJ}
\left. \frac{dE_{\textrm{wave}}}{dt} \right|_{J_{ij}}&=&
\frac{32}{5} \left(\frac{\mu}{M}\right)^2 v^{10} \times{}\nn\\
& &{}\times\left[ \frac{v^2}{36}-\frac{S_1 v^3}{12 M^2}+\frac{S_1^2 v^4}{16 M^4}+{\mathcal O}(v^5) \right],
\eea
see Ref.~\cite{Ryan}. We should also put by hand further additional contributions of $dE_{\textrm{wave}}/dt$,
due to post-newtonian corrections. The corresponding contributions, up to $v^4$,
are simply numerical if one computes them from perturbation analysis in a Schwarzschild background.
A comparison though, between
the final formula (55) of \cite{Ryan} for $dE/dt$, and formula (3.13) of \cite{Shibetal}, which is
based on perturbations on a Kerr background, shows that at least up to $v^4$ order, the terms of
\cite{Ryan} that include $S_1$ and $M_2$, which come from the corresponding contributions
that are given by Eqs.~(\ref{dEdtI},\ref{dEdtJ}),  in the case of a Kerr metric, are equal to the ones of Ref. \cite{Shibetal}.
This agreement indicates
that up to $v^4$ order we can simply add the following numerical post-newtonian terms (the corresponding
terms of \cite{Shibetal} if we set $q=0$) in the rest contributions of $dE_{\textrm{wave}}/dt$
\bea
\label{dEdtPN}
\left. \frac{dE_{\textrm{wave}}}{dt} \right|_{PN}&=&
\frac{32}{5} \left(\frac{\mu}{M}\right)^2 v^{10} \times
\left[ -\frac{1247}{336}v^2 +{}\right.\nn\\
& &{}+\left. 4\pi v^3 - \frac{44711}{9072} v^4 +{\mathcal O}(v^5) \right].
\eea
Finally, in order to compute the number of cycles $\Delta N$, one has to add up all contributions of
 $dE_{\textrm{wave}}/dt$ and combine them with the expression for $\Delta E/\mu$,
 which is given above (see Eq.~(\ref{delE})).

To get expressions for all these quantities, that are
straightforwardly connected to the moments of the central object, and  can be observationally measured, 
first one has to reexpress the metric
functions in terms of all moments. Also the relation between the radius $\rho$ and the orbital
frequency of the test body $\Omega$, through moments, is necessary so as to finally express the four
measurable quantities as power series of $v$.

%%%%%%%%%%%%%%%%%%%%%
\subsection{\label{sec:2.2}
Moments describing spacetime}
%%%%%%%%%%%%%%%%%%%%%
In our paper, we consider only stationary axisymmetric objects that are symmetric
with respect to their equatorial plane. These symmetries are more or less realistic assumptions
for a quiescent massive rotating astrophysical body around which much smaller bodies orbit.
The metric of such a central object alone could be written in $(t,\rho, z, \phi)$ coordinates, in the form of
Papapetrou metric \cite{pap}.
\be
\label{papmet}
ds^2=-F(dt-\omega ~d\phi)^2+\frac{1}{F} \left[
e^{2\gamma} (d\rho^2 + dz^2) + \rho^2 d\phi^2 \right],
\ee
where $F,\omega$, and $\gamma$ are the three functions that fully determine a specific metric. These
are functions of $\rho$ and $|z|$ only, due to axisymmetry and reflection symmetry. Einstein's equations in vacuum
guarantee that once $F$ and $\omega$ are given, $\gamma$ can be easily computed (see \cite{Wald}).
Once we incorporate an electromagnetic field in the vacuum around the compact object, the source
of which is the compact object itself,
which allows spacetime to have the same symmetries, the metric above still describes the
electrovacuum  spacetime, but now the metric and the electromagnetic field should satisfy the
Einstein-Maxwell equations. In order to fully compute the
metric functions one more complex function, $\Phi$,
which is related to the electromagnetic field, is necessary.
$F$, $\omega$, and $\Phi$ themselves can be determined by solving the so-called Ernst equations
\cite{ernst1,ernst2}, which are nothing more than the Einstein-Maxwell equations in a different form.
It is a system of non-linear complex differential equations of second order:
\bea
\label{EinstE}
(\Re ({\mathcal E}) + |\Phi|^2 ) \nabla^2 {\mathcal E} &=&
(\bn {\mathcal E} + 2 \Phi^\ast \bn \Phi )
\cdot \bn {\mathcal E}, \\
\label{EinstF}
(\Re ({\mathcal E}) + |\Phi|^2 ) \nabla^2 {\Phi} &=&
(\bn {\mathcal E} + 2 \Phi^\ast \bn \Phi )
\cdot \bn {\Phi},
\eea
where $\bn$ denotes the gradient in a cartesian 3D space $(\rho,z,\phi)$ and $\Re(\ldots),\Im(\ldots)$, here
and henceforth, denote the real and imaginary
part, respectively, of the complex function in parentheses.
An asterisc $^\ast$ denotes complex conjugate.
The third metric function, $\gamma$, is then easily computed by integrating the partial derivatives
$\partial \gamma/\partial \rho$, $\partial \gamma/\partial z$, which are given as functions of
derivatives of all other functions
\bea
\label{gammaeq}
\frac{\partial \gamma}{\partial \rho}&=&\frac{1}{4}\frac{\rho}{g_{tt}^{2}}\left[\left(\frac{\partial g_{tt}}{\partial \rho}\right)^{2}-\left(\frac{\partial g_{tt}}{\partial z}\right)^{2}\right]-{}\nn\\
& &{}-\frac{1}{4}\frac{g_{tt}^{2}}{\rho}\left[\left(\frac{\partial (g_{t\phi}/g_{tt})}{\partial \rho}\right)^{2}-\left(\frac{\partial (g_{t\phi}/g_{tt})}{\partial z}\right)^{2}\right]-{}\nn\\
& &{}-\frac{\rho}{g_{tt}}\left[\left(\frac{\partial \Re(\Phi)}{\partial \rho}\right)^{2}-\left(\frac{\partial \Re(\Phi)}{\partial z}\right)^{2}\right]-{}\nn\\
& &{}-\frac{\rho}{g_{tt}}\left[\left(\frac{\partial \Im(\Phi)}{\partial \rho}\right)^{2}-\left(\frac{\partial \Im(\Phi)}{\partial z}\right)^{2}\right],
\eea
\bea
\label{gammaeqz}
\frac{\partial \gamma}{\partial z}&=&\frac{1}{2}\frac{\rho}{g_{tt}^{2}}\frac{\partial g_{tt}}{\partial \rho}\frac{\partial g_{tt}}{\partial z}-\frac{1}{2}\frac{g_{tt}^{2}}{\rho}\frac{\partial (g_{t\phi}/g_{tt})}{\partial \rho}\frac{\partial (g_{t\phi}/g_{tt})}{\partial z}-{}\nn\\
& &{}-2\frac{\rho}{g_{tt}}\frac{\partial \Re(\Phi)}{\partial \rho}\frac{\partial \Re(\Phi)}{\partial z}-2\frac{\rho}{g_{tt}}\frac{\partial \Im(\Phi)}{\partial \rho}\frac{\partial \Im(\Phi)}{\partial z}.
\eea

The relation of the two complex functions, $\mathcal E$ and $\Phi$, with the metric functions is the following
\bea
\label{ernstpot}
\mathcal{E}&=&(F-\left|\Phi\right|^2)+i\varphi,
%=\frac{\sqrt{\rho^{2}+z^{2}}-\tilde{\xi}}{\sqrt{\rho^{2}+z^{2}}+\tilde{\xi}}\nn\\
%\Phi&=&\frac{\tilde{q}}{\sqrt{\rho^{2}+z^{2}}+\tilde{\xi}}
\eea
%%%%%%%%%%
where $\varphi$ is related with $g_{t\phi}$ through
%%%%%%%%%
\bea
\label{intgtf}
g_{t\phi}=F \omega&=& F \int_{\rho}^{\infty} \!\!d\rho'
\frac{\rho'}{F^{2}}\left(\frac{\partial\varphi}{\partial z}+
2{\Re}(\Phi)\frac{\partial{\Im}(\Phi)}{\partial z}-{}\right.\nn\\
& &{}-\left.\left.
2{\Im}(\Phi)\frac{\partial{\Re}(\Phi)}{\partial z}
\right)\right|_{z=\textrm{const}}.
\eea
%%%%%%%%%
Note that there is a sign difference in Eq.~(22) of \cite{Ryan}, which has been corrected in
a later paper of Ryan \cite{RyanCORRECT}, and comes from an odd convention of $\omega$
used by Ernst (see relevant comment of \cite{israel}).

Instead of $\mathcal E$ and $\Phi$, one could use two new complex functions $\tilde \xi$ and $\tilde q$,
that play the role of gravitational potential and Coulomb potential respectively, and are more directly
connected to the mass and electromagnetic moments of the central body. These potentials are related
to the Ernst functions by
\bea
\label{etoxi}
\mathcal{E} &=& \frac{\sqrt{\rho^{2}+z^{2}}-\tilde{\xi}}{\sqrt{\rho^{2}+z^{2}}+\tilde{\xi}} \\
\Phi&=&\frac{\tilde{q}}{\sqrt{\rho^{2}+z^{2}}+\tilde{\xi}},
\eea
and can be written as power series expansions at infinity
\bea
\label{expxiq}
\tilde {\xi } = \sum\limits_{i,j = 0}^\infty {a_{ij}\bar {\rho }^i\bar {z}^j},\qquad
\tilde {q} = \sum\limits_{i,j = 0}^\infty {b_{ij}\bar {\rho }^i\bar {z}^j},
\eea
%%%%%%%%%%%%
where
\be
\label{barred}
\bar \rho \equiv \frac{\rho}{\rho^2+z^2},\qquad \bar z \equiv \frac{z}{\rho^2+z^2},
\ee
and $a_{ij},b_{ij}$ are coefficients that vanish when $i$ is odd.
This reflects the analyticity of the potentials on the $z$-axis. The tilded quantities, here and henceforth,
are the conformally transformed ones, which are essential for calculating the moments (see \cite{Geroch}).

Due to Ernst equations (\ref{EinstE},\ref{EinstF})
the above power expansion coefficients, $a_{ij}$ and $b_{ij}$, are interrelated through the
following complicated recursive relations
\bea
\label{abrecur1}
 \left(r+2\right)^2a_{r+2,s}&=&-(s+2)(s+1)a_{r,s+2}+{} \nn\\
 &+& \sum_{k,l,m,,n,p,g}
(a_{kl} a_{mn}^\ast-b_{kl}b_{mn}^\ast) \times{} \nn\\
&& \big[
    a_{pg} (p^2 + g^2 - 4p - 5g - 2pk - 2gl -2)  {} \nn\\
&+& a_{p + 2,g - 2} (p + 2)(p + 2 - 2k)  {}\nn\\
&+& a_{p - 2,g + 2} (g + 2)(g + 1 - 2l) \big],
\eea
%%%%%%%%%%%%%
and
%%%%%%%%%%%%
\bea
\label{abrecur2}
 \left(r+2\right)^2b_{r+2,s}&=&-(s+2)(s+1)b_{r,s+2}+{} \nn\\
&+& \sum_{k,l,m,n,p,g}
(a_{kl} a_{mn}^\ast-b_{kl}b_{mn}^\ast)\times {} \nn\\
&& \big[
    b_{pg} (p^2 + g^2 - 4p - 5g - 2pk - 2gl -2) {} \nn\\
&+& b_{p + 2,g - 2} (p + 2)(p + 2 - 2k) {}\nn\\
&+& b_{p - 2,g + 2} (g + 2)(g + 1 -2l)
\big],
\eea
%%%%%%%%%%%%%
where $m=r-k-p$ , $0\leq k \leq r$, $0 \leq p \leq r-k$ , with $k$ and $p$ even,
and $n=s-l-g$, $0 \leq l \leq s+1$ , and $-1 \leq g \leq s-l$.
Essentially, these relations are simply an algebraic version of Einstein-Maxwell equations
for the coefficients of the power expansion of the metric and the electromagnetic field tensor.
The recursive relations (\ref{abrecur1},\ref{abrecur2}) could be used to build
the whole power series of $\tilde{\xi}$
and $\tilde{q}$ from a full knowledge of the metric on the axis of symmetry
\bea
\tilde{\xi}(\bar\rho=0)=\sum_{i=0}^{\infty} m_i {\bar z}^i ,\qquad
\tilde{q}(\bar\rho=0)=\sum_{i=0}^{\infty} q_i {\bar z}^i.
\eea

In \cite{SotiApos} a method of calculating the complex multipole moments of the central object in terms of
the $m_{i}$'s and $q_{i}$'s is presented. In brief, the gravitational moments are given by
%%%%%%%%%%
\begin{equation}
\label{pnfinal}
P_n = \frac{1}{(2n - 1)!!}S_0^{(n)},
\end{equation}
%%%%%%%%%%%%
where $S_{a}^{(n)}$ are computed recursively by
%%%%%%%%%%%%%
\bea
\label{mms}
S_0^{(0)}&=&\tilde {\xi },\nn\\
S_0^{(1)}&=&\frac{\partial }{\partial {\bar z}}S_0^{(0)},\nn\\
S_1^{(1)}&=&\frac{\partial }{\partial {\bar \rho} }S_0^{(0)},\nn\\
S_a^{(n)}&=&\frac{1}{n}\Bigg\{a\frac{\partial }{\partial {\bar \rho} }S_{a
- 1}^{(n - 1)} + (n - a)\frac{\partial }{\partial {\bar z}}S_a^{(n - 1)} +{}\nn\\
& &{}+ a\left[
(a + 1 - 2n)\gamma _1 - \frac{a - 1}{{\bar \rho} } \right]S_{a - 1}^{(n - 1)}+{}\nn\\
& &{}+ (a - n)(a + n - 1)\gamma _2 S_a^{(n - 1)} +{}\nn\\
& & {}+ a(a - 1)\gamma _2 S_{a -2}^{(n - 1)} +{}\nn\\
& &{}+ (n - a)(n - a - 1)\left( \gamma _1 - \frac{1}{{\bar \rho} } \right)S_{a +
1}^{(n - 1)} -{}\nn\\
& &{}- \left( n - \frac{3}{2} \right)\left[ a(a - 1)\tilde{R}_{11} S_{a - 2}^{(n - 2)} +\right.{}\nn\\
& &{}+ 2a(n - a)\tilde {R}_{12} S_{a - 1}^{(n - 2)} +{}\nn\\
& &{}+\left.(n -a)(n - a - 1)\tilde {R}_{22} S_a^{(n - 2)}\right]\Bigg\}.
\eea

%%%%%%%%%%%%%
The electromagnetic moments $Q_{n}$ are computed from exactly the same formulae, by simply
replacing the initial term $S_{0}^{(0)}$ with $\tilde{q}$ instead of $\tilde{\xi}$.
$\tilde{R}_{11}$, $\tilde{R}_{12}$ and $\tilde{R}_{22}$ are given by
%%%%%%%%%%
\bea
\tilde{R}_{ij}&=&(\bar{r}^{2}\tilde{\xi}^{\ast}\tilde{\xi}-
                  \bar{r}^{2}\tilde{ q }^{\ast}\tilde{ q }-1)^{-2}
                  (D_{i}\tilde{\xi}D_{j}\tilde{\xi}^{\ast}+D_{i}\tilde{\xi}^{\ast}D_{j}\tilde{\xi}-{}\nn\\
& & {}-D_{i}\tilde{q}D_{j}\tilde{q}^{\ast}-D_{i}\tilde{q}^{\ast}D_{j}\tilde{q}+\tilde{s}_{i}\tilde{s}_{k}^{\ast}+\tilde{s}_{i}^{\ast}\tilde{s}_{k}),\nn\\
\eea
%%%%%%%%%%
where
%%%%%%%
\bea
{\bar r}^2&=&{\bar \rho}^2+{\bar z}^2,\nn \\
D_{1}&=&\bar{z}\frac{\partial}{\partial \bar{\rho}}-\bar{\rho}\frac{\partial}{\partial\bar{z}},\nn\\
D_{2}&=&\bar{\rho}\frac{\partial}{\partial \bar{\rho}}+\bar{z}\frac{\partial}{\partial\bar{z}}+1,\nn\\
\tilde{s}_{i}&=&\bar{r}\tilde{\xi}D_{i}\tilde{q}-\bar{r}\tilde{q}D_{i}\tilde{\xi}.
\eea
%%%%%%%%
and $\gamma_{1}\equiv\gamma_{,\rho}$ and $\gamma_{2}\equiv\gamma_{,z}$ can be expressed in term of $\tilde{R}_{ij}$ as
%%%%%%%%
\bea
\label{gammaxiq}
\gamma_{1}&=&\frac{1}{2}\bar{\rho}(\tilde{R}_{11}-\tilde{R}_{22}),\nn\\
\gamma_{2}&=&\bar{\rho}\tilde{R}_{12}.
\eea
%%%%%%%
The mass moments $M_{n}$ and the mass-current moments $S_{n}$ are related to $P_{n}$ by
%%%%%%%%%%%%
\be
\label{massangular}
P_{n}=M_{n}+i S_{n},
\ee
%%%%%%%%%%%%
whereas the electric  moments $E_{n}$ and the magnetic  moments $H_{n}$ are related to $Q_{n}$ by
%%%%%%%
\be
\label{electricmagnetic}
Q_{n}=E_{n}+i H_{n}.
\ee
%%%%%%%

Since this algorithm can be used to evaluate the moments in terms of the $m_{i}$'s and $q_{i}$'s,
one can invert these relations and express the $m_{i}$'s and $q_{i}$'s in terms of the moments:
%%%%%%%%
\bea
\label{lom}
m_{n}&=&a_{0n}=M_{n}+iS_{n}+\textrm{LOM},\nn\\
q_{n}&=&b_{0n}=E_{n}+iH_{n}+\textrm{LOM},
\eea
%%%%%%%%
where ``LOM'' stands for lower order multipole moments of any type.
Thus, we can use the recursive relations (\ref{abrecur1})
 and (\ref{abrecur2}) to evaluate the $a_{ij}$ and $b_{ij}$ coefficients in terms of the moments.
Finally, following the procedure
presented in the beginning of this subsection (Eqs.~(\ref{gammaeq}-\ref{expxiq}))
we can express the metric functions, and their first and second derivatives
as power series of $\rho$ and $z$ with coefficients that are simple algebraic functions of the moments
of the massive body. Since in our study we have confined the motion of the test particle
on the equatorial plane, we  actually need to compute everything at $z=0$ which makes
calculations far simpler than what they seem.

%%%%%%%%%%%%%%%%%%%%%%%%%
\subsection{\label{sec:2.3}
Reflection symmetry and electromagnetic moments}
%%%%%%%%%%%%%%%%%%%%%%%%%
Following Ernst \cite{ernst1,ernst2} and using Papapetrou's metric
(\ref{papmet}) we end up with the following Einstein-Maxwell equations:
\begin{gather}
\label{elfield1}
\qquad\bn\cdot\left[\rho^{-2}F\left(\bn A_{3}-\omega\bn A_{4}\right)\right]=0,\\
\label{elfield2}
\bn\cdot\left[F^{-1}\bn A_{4}+\rho^{-2}F\omega\left(\bn A_{3}-\omega\bn A_{4}\right)\right]=0,\\
\label{elfield3}
\bn\cdot\left[\rho^{-2}F^{2}\bn\omega-4\rho^{-2}FA_{4}\left(\bn A_{3}-\omega\bn A_{4}\right)\right]=0,\\
\label{elfield4}
F\nabla^{2}F=\bn F\cdot\bn F-\rho^{-2}f^{4}\bn\omega\cdot\bn\omega+2F\bn A_{4}\cdot\bn A_{4}+{}\nn\\
{}+2\rho^{2}F^{3}\left(\bn A_{3}-\omega\bn A_{4}\right)\cdot\left(\bn A_{3}-\omega\bn A_{4}\right),
\end{gather}
%%%%%%%%%%%%%%%%%%%%%%%%%%%%%%%%%%%%%%
where $A_{3}$ and $A_{4}$ denote the $A_{\phi}$ and $A_{t}$ components of the electromagnetic 4-potential respectively,
and $\bn$ is the three dimensional divergence operator in Weyl coordinates. As we have
already mentioned, the three metric functions $F$, $\omega$ and $\gamma$ are functions of $\rho$
and $\left|z\right|$ only, due to the assumed symmetries. From the equations above one cannot
easily tell what are the symmetries, if any, that are inherited in
$A_{3}$ and $A_{4}$. It is obvious though that $A_{3}$ and $A_{4}$ being either both odd
or both even functions of $z$ is consistent with the reflection symmetry of the metric functions.
We will argue that these are the only reasonable types of the electromagnetic field.

The Ernst potential $\Phi$ is defined by Ernst through the following relations:
%%%%%%%%%%%%%%%%%%%%%%%%%%%%%%%%%%%%%
\be
\label{scalmag}
\rho^{-1}f\left(\bn A_{3}-\omega\bn A_{4}\right)=\hat{n}\times\bn A_{3}',
\ee
\be
\label{ernfi}
\Phi=A_{4}+iA_{3}'.
\ee
where $\hat{n}$ is the unit vector in the azimouthal direction.
From Eq.~(\ref{intgtf}), we obtain the metric function $g_{t\phi}$.  Since we want to end up with
metric functions that are even functions of $z$, then the whole intergrand should be even as well.
This could be acomplished if $\varphi$ is an odd function of $z$ and the real and imaginary part of
$\Phi$ are either even and odd or odd and even functions of $z$ respectively. If none of the above holds
then in order to get an even function for $g_{t\phi}$, one must impose a restraining functional
relation between $\varphi$ and $\Phi$. But $\varphi$ and $\Phi$ should be independent in order to
describe a generic spacetime with the symmetries mentioned above.
Therefore, by virtue of Eqs.~(\ref{scalmag},\ref{ernfi}), either $A_4$ and $A_3$ are both even, or both odd functions of $z$.
Consequently, for both cases
the $\Re({\mathcal E})$ is an even and the $\Im({\mathcal E})$ is an odd function of $z$.

Now, we can use this information to conclude that, by virtue of Eq.~(\ref{etoxi}), the action of reflection
symmetry leaves $\Re(\tilde{\xi})$ invariant and changes the sign of $\Im(\tilde{\xi})$.
Thus only even order mass moment and odd order current mass moments will occur \cite{hansennote}.
For the electromagnetic moments things are not univocal since we have
two discernible electromagnetic cases that are consistent with the symmetries of
the metric. If both $A_{3}$ and $A_{4}$ are
odd functions of $z$ then the  action of the reflection symmetry
leaves $\Re(\tilde{q})$ invariant and reverses the sign of $\Im(\tilde{q})$, which means
that only even order electric field moments and odd order magnetic field moments will occur. On the other
hand if both $A_{3}$ and $A_{4}$ are even functions of $z$ then the  action of the reflection symmetry
leaves $\Im(\tilde{q})$ invariant, but reverses the sign of $\Re(\tilde{q})$, which means
that in this case only odd order electric field moments and even order magnetic field moments will occur.

%%%%%%%%%%%%%%%%%%%%
%%%%%%%%%%%%%%%%%%%%
\section{\label{sec:3}The power expansion formulae}
%%%%%%%
%%%%%%%%%%%%%%%%%%%%

Combining all formulae that are given in Sec.~\ref{sec:2} eventually we can express all four
measurable quantities as power series of $v \equiv (M \Omega)^{1/3}$ with coefficients that
have explicit dependence on all four types of moments. The choice of $v$ as a dimensionless parameter
to expand all physical quantities is warranted from the fact that the inspiral phase of a binary,
the best exploitable part in gravitational-wave analysis \cite{CutlFlan}, involves
comparatively low magnitudes of $v$.
All measurable quantities have been transformed to a dimensionless form as well, for example by dividing
the two frequencies $\Omega_\rho$, $\Omega_z$, by the orbital frequency $\Omega$.

Since the metric functions and their derivatives, are expressed as functions only of $\rho$ at the equatorial
plane, in order to express all measurable quantities as power series of $v$, we need also a
power series expansion of $\rho$ with respect to $v$, or equivalently $\Omega$. Thus we have to
invert the function $\Omega(\rho)$, at least as a power expansion. From an elementary analysis
of circular geodesics on the equatorial plane (see \cite{Ryan}) we know that
\be
\label{Omega}
\Omega=\frac{ -g_{t \phi,\rho} +
\sqrt{(g_{t \phi,\rho})^2-(g_{tt,\rho})(g_{\phi \phi,\rho})} }{ g_{\phi \phi,\rho} } .
\ee

In the following part of this section we explain the algorithm that one should follow,
in order to obtain the power series for
$\Omega_\rho/\Omega$, $\Omega_z/\Omega$, $\Delta E/\mu$, and $\Delta N$. One starts with a power series of
$\tilde{\xi}$ and $\tilde{q}$ of the form given by Eq.~(\ref{expxiq}). Since no higher than second
derivatives of the metric functions with respect to $z$ are necessary, one should keep $a_{ij}$'s
and $b_{ij}$'s with $0 \leq j \leq 2$, and as many values of $i$ as one needs to carry the power
series expansion of the measurable quantities at a desirable order. In our paper where all quantities are
written up to no higher than $v^{11}$ order, we only need $i$'s in the interval $0\leq i \leq 4$. All quantities
that are expressed as power series of $\bar{\rho}$ and $\bar{z}$, are evaluated at ${\bar z}=z=0$
at the end, and thus, all expressions are finally power series of ${\bar \rho}=1/\rho$, 
due to Eq.~(\ref{barred}). Although
the $a_{ij}$ and $b_{ij}$ are polynomials of various moments, from the practical point of view
it is preferable to keep them as they are, and replace them by their moments
dependence only at the final expressions.
Then from $\tilde{\xi}$ and $\tilde{q}$ we construct ${\mathcal E}$, $\Phi$, and $F$, $\varphi$
(cf., Eqs.~(\ref{ernstpot},\ref{etoxi})). These are sufficient to build all metric functions
through Eqs.~(\ref{gammaeq},\ref{gammaeqz},\ref{intgtf}). Next, following the procedure described above,
we expand $\Omega$ as a power series of $1/\rho$, by virtue of Eq.~(\ref{Omega}). This series
is inverted and in this way we obtain $1/\rho$ as a power series of $\Omega$, which
then can easily be turned into a power series of the dimensionless parameter $v$.

Now, the power series representing $1/\rho$ will replace all $1/\rho$ terms appearing at
the expansions of the metric, its derivatives, and all other physical quantities depending on them
(Eqs.~(\ref{omegas}-\ref{dEdtI})).  Finally, one has to rewrite
the $a_{ij}$ and $b_{ij}$ terms appearing at the coefficients of all these power series
as polynomials of the various moments. The recursive relations (\ref{abrecur1},\ref{abrecur2})
relate all $a_{ij}$ and $b_{ij}$ with  $m_k \equiv a_{0k}$ and $q_k \equiv b_{0k}$,
which are directly related to the scalar moments of spacetime through Eqs.~(24,25) of Ref.~\cite{SotiApos}.

The algorithm described in the previous two paragraphs has been carried out with Mathematica,
and has been checked for the following two subcases: (i) When all electromagnetic fields are
turned off, by erasing all electromagnetic moments ($E_l=H_l=0$), our expressions for
$\Omega_\rho$, $\Omega_z$, $\Delta E/\mu$, $\Delta N$ are identical to the ones computed by
Ryan \cite{Ryan}. (ii) For the Kerr-Newman metric it is quite easy to compute $\Omega$,
$\Omega_\rho$, $\Omega_z$, and $\Delta E/\mu$ for a quasi-equatorial, quasi-circular orbit.
Actually, there is no need to use Weyl coordinates to describe the metric; one could
simply work with the metric in the usual Boyer-Lindquist coordinates (see Eq.~(33.2) of \cite{MTW}),
and compute everything according to the formulae given above, by replacing the derivatives
with respect to $z$, with the corresponding derivatives with respect to $\theta$
around $\theta=\pi/2$. The expressions for $\rho$ though should be replaced with
$(g^2_{t \phi}-g_{tt} g_{\phi \phi})^{1/2}$. Thus, if in the power series for
$\Omega_\rho$, $\Omega_z$, and $\Delta E/\mu$ that are written below (cf., Eqs.~(\ref{wpprosw},
\ref{wzprosw},\ref{deprosm})), one makes the following
substitutions for the moments
\bea
M_{2l}&=&(-1)^l M a^{2l},\quad S_{2l+1}=(-1)^l M a^{2l+1},\nn\\
E_{2l}&=&(Q/M) M_{2l},\quad H_{2l+1}=(Q/M) S_{2l+1},
\eea
according to \cite{SotiApos}, the expressions we  obtain are identical to the ones obtained directly
from the Kerr-Newman metric.

There is one more thing that should be pointed out before we write down the power series
for all four observable quantities. As is explained in Sec.~\ref{sec:2.3} there are two possible
cases for the electromagnetic field that lead to reflection-symmetric spacetimes.
The first case (with odd $A_3$ and $A_4$ as functions of $z$) is the one that describes
an electric field that is reflection symmetric (like in a monopole electric field), and a magnetic
field that is reflection antisymmetric (like in a magnetic dipole field). Henceforth we shall call
this case electric-symmetric case ($es$). In that case only the even electric moments and the odd
magnetic moments show up in the moment analysis of spacetime. Thus $b_{0l}$ is
real for even $l$'s and purely imaginary for odd $l$'s.
The other case (with even $A_3$ and $A_4$ as functions of $z$) is the one that describes
an electric field that is reflection antisymmetric (like in a dipole electric field), and a magnetic
field that is reflection symmetric (like in a magnetic quadrupole field).
Henceforth we shall call
this case magnetic-symmetric case ($ms$). In that case only the odd electric moments and the even
magnetic moments show up in the moment analysis of spacetime. Thus $b_{0l}$ is
real for odd $l$'s and purely imaginary for even $l$'s. Although classically we do not expect
the central object to carry any magnetic monopole, the zeroth order magnetic moment
shows up formally in the terms of a generic magnetic-symmetric case, and thus we have not omitted it.

The power series expansion for $\Omega_\rho$, $\Omega_z$, and $\Delta E/\mu$ take the following form:
\bea
\label{wpprosw}
\frac{\Omega_{\rho}}{\Omega}&=&\sum\limits_{n=2}^{\infty}R_{n}v^{n},\\
\label{wzprosw}
\frac{\Omega_{z}}{\Omega}&=&\sum\limits_{n=3}^{\infty}Z_{n}v^{n},\\
\label{deprosm} 
\frac{\Delta E}{\mu}&=&\sum\limits_{n=2}^{\infty}A_{n}v^{n}, 
\eea 
while the corresponding
coefficients, up to 9nth order for the two frequencies and up to 11nth order for the radiated energy, in the two
distinct electromagnetic cases ($es$) and ($ms$) are
\begin{widetext}
%%%%%%%%%%%%%%
\bea
\label{wpprosw1}
R^{(es)}_{2}&=&\left(3-\frac{1}{2}\frac{E^{2}}{M^{2}}\right),\quad
R^{(es)}_{3}=-4\frac{S_{1}}{M^{2}},\nn\\
R^{(es)}_{4}&=&\frac{9}{2}-\frac{3}{2}\frac{M_{2}}{M^{3}}-2\frac{E^{2}}{M^{2}}-\frac{13}{24}\frac{E^{4}}{M^{4}},\quad
R^{(es)}_{5}=-10\frac{S_{1}}{M^{2}}-\frac{10}{3}\frac{S_{1}E^{2}}{M^{4}}+5\frac{H_{1}E}{M^{3}},\nn\\
R^{(es)}_{6}&=&\frac{27}{2}-2\frac{S_{1}^{2}}{M^{4}}-\frac{21}{2}\frac{M_{2}}{M^{3}}-
\frac{33}{4}\frac{E^{2}}{M^{2}}-\frac{1}{8}\frac{E^{4}}{M^{4}}-
\frac{35}{48}\frac{E^{6}}{M^{6}}-\frac{11}{4}\frac{M_{2}E^{2}}{M^{5}}+3\frac{E_{2}E}{M^{4}},\nn\\
R^{(es)}_{7}&=&-48\frac{S_{1}}{M^{2}}-5\frac{S_{1}M_{2}}{M^{5}}+9\frac{S_{3}}{M^{4}}-
\frac{5}{3}\frac{S_{1}E^{2}}{M^{4}}-\frac{91}{18}\frac{S_{1}E^{4}}{M^{6}}+
19\frac{H_{1}E}{M^{3}}+\frac{11}{2}\frac{H_{1}E^{3}}{M^{5}},\nn\\
R^{(es)}_{8}&=&\frac{405}{8}+\frac{2243}{84}\frac{S_{1}^{2}}{M^{4}}-
\frac{661}{14}\frac{M_{2}}{M^{3}}-\frac{21}{8}\frac{M_{2}^{2}}{M^{6}}+
\frac{15}{4}\frac{M_{4}}{M^{5}}-\frac{81}{2}\frac{E^{2}}{M^{2}}+\frac{53}{16}\frac{E^{4}}{M^{4}}-
\frac{19}{216}\frac{E^{6}}{M^{6}}-\frac{11339}{10368}\frac{E^{8}}{M^{8}}-
\frac{251}{21}\frac{M_{2}E^{2}}{M^{5}}-{}\nn\\
& &{}-\frac{85}{16}\frac{M_{2}E^{4}}{M^{7}}+
\frac{75}{4}\frac{E_{2}E}{M^{4}}+
\frac{35}{6}\frac{E_{2}E^{3}}{M^{6}}-\frac{29}{28}\frac{H_{1}^{2}}{M^{4}}-\frac{103}{9}\frac{S_{1}^{2}E^{2}}{M^{6}}+\frac{13}{3}\frac{H_{1}S_{1}E}{M^{5}},\nn\\
R^{(es)}_{9}&=&-243\frac{S_{1}}{M^{2}}-16\frac{S_{1}^{3}}{M^{6}}+4\frac{S_{1}M_{2}}{M^{5}}+
45\frac{S_{3}}{M^{4}}+54\frac{S_{1}E^{2}}{M^{4}}-\frac{125}{12}\frac{S_{1}E^{4}}{M^{6}}-
\frac{35}{4}\frac{S_{1}E^{6}}{M^{8}}-\frac{41}{2}\frac{S_{1}M_{2}E^{2}}{M^{7}}+
8\frac{S_{1}E_{2}E}{M^{6}}+{}\nn\\
& &{}+\frac{165}{2}\frac{H_{1}E}{M^{3}}+\frac{20}{3}\frac{H_{1}E^{3}}{M^{5}}+
\frac{205}{24}\frac{H_{1}E^{5}}{M^{7}}+\frac{15}{2}\frac{H_{1}M_{2}E}{M^{6}}-
\frac{7}{2}\frac{H_{1}E_{2}}{M^{5}}-\frac{21}{2}\frac{H_{3}E}{M^{5}}+\frac{25}{2}\frac{S_{3}E^{2}}{M^{4}},\nn
\eea
%%%%%%%%%%%%%%%%%%%%
\bea
\label{wpprosw2}
R^{(ms)}_{2}&=&\left(3-\frac{1}{2}\frac{H^{2}}{M^{2}}\right),\quad
R^{(ms)}_{3}=-4\frac{S_{1}}{M^{2}},\nn \\
R^{(ms)}_{4}&=&\frac{9}{2}-\frac{3}{2}\frac{M_{2}}{M^{3}}-2\frac{H^{2}}{M^{2}}-
\frac{13}{24}\frac{H^{4}}{M^{4}},\quad
R^{(ms)}_{5}=-10\frac{S_{1}}{M^{2}}-\frac{10}{3}\frac{S_{1}H^{2}}{M^{4}}-5\frac{E_{1}H}{M^{3}},\nn\\
R^{(ms)}_{6}&=&\frac{27}{2}-2\frac{S_{1}^{2}}{M^{4}}-\frac{21}{2}\frac{M_{2}}{M^{3}}-
\frac{33}{4}\frac{H^{2}}{M^{2}}-\frac{1}{8}\frac{H^{4}}{M^{4}}-\frac{35}{48}\frac{H^{6}}{M^{6}}-
\frac{11}{4}\frac{M_{2}H^{2}}{M^{5}}+3\frac{H_{2}H}{M^{4}},\nn\\
R^{(ms)}_{7}&=&-48\frac{S_{1}}{M^{2}}-5\frac{S_{1}M_{2}}{M^{5}}+9\frac{S_{3}}{M^{4}}-
\frac{5}{3}\frac{S_{1}H^{2}}{M^{4}}-\frac{91}{18}\frac{S_{1}H^{4}}{M^{6}}-19\frac{E_{1}H}{M^{3}}-
\frac{11}{2}\frac{E_{1}H^{3}}{M^{5}},\nn\\
R^{(ms)}_{8}&=&\frac{405}{8}+\frac{2243}{84}\frac{S_{1}^{2}}{M^{4}}-\frac{661}{14}\frac{M_{2}}{M^{3}}-
\frac{21}{8}\frac{M_{2}^{2}}{M^{6}}+\frac{15}{4}\frac{M_{4}}{M^{5}}-\frac{81}{2}\frac{H^{2}}{M^{2}}+
\frac{53}{16}\frac{H^{4}}{M^{4}}+
\frac{19}{216}\frac{H^{6}}{M^{6}}-\frac{11339}{10368}\frac{H^{8}}{M^{8}}-
\frac{251}{21}\frac{M_{2}H^{2}}{M^{5}}-{}\nn\\
& &{}-\frac{85}{16}\frac{M_{2}H^{4}}{M^{7}}+
\frac{75}{4}\frac{H_{2}H}{M^{4}}+
\frac{35}{6}\frac{H_{2}H^{3}}{M^{6}}-\frac{29}{28}\frac{E_{1}^{2}}{M^{4}}-
\frac{103}{9}\frac{S_{1}^{2}H^{2}}{M^{6}}-\frac{13}{3}\frac{E_{1}S_{1}H}{M^{5}},\nn\\
R^{(ms)}_{9}&=&-243\frac{S_{1}}{M^{2}}-16\frac{S_{1}^{3}}{M^{6}}+4\frac{S_{1}M_{2}}{M^{5}}+
45\frac{S_{3}}{M^{4}}+54\frac{S_{1}H^{2}}{M^{4}}-\frac{125}{12}\frac{S_{1}H^{4}}{M^{6}}-
\frac{35}{4}\frac{S_{1}H^{6}}{M^{8}}-\frac{41}{2}\frac{S_{1}M_{2}H^{2}}{M^{7}}+
8\frac{S_{1}H_{2}H}{M^{6}}-{}\nn\\
& &{}-\frac{165}{2}\frac{E_{1}H}{M^{3}}-\frac{20}{3}\frac{E_{1}H^{3}}{M^{5}}-
\frac{205}{24}\frac{E_{1}H^{5}}{M^{7}}-\frac{15}{2}\frac{E_{1}M_{2}H}{M^{6}}+
\frac{7}{2}\frac{E_{1}H_{2}}{M^{5}}+\frac{21}{2}\frac{E_{3}H}{M^{5}}+\frac{25}{2}\frac{S_{3}H^{2}}{M^{4}},\nn
\eea
%%%%%%%%%%%
%%%%%%%%%%%
\bea
\label{wzprosw1}
Z^{(es)}_{3}&=&2\frac{S_{1}}{M^{2}},\quad
Z^{(es)}_{4}=\frac{3}{2}\frac{M_{2}}{M^{3}},\quad
Z^{(es)}_{5}=-\frac{H_{1}E}{M^{3}}+2\frac{S_{1}E^{2}}{M^{4}},\nn\\
Z^{(es)}_{6}&=&7\frac{S_{1}^{2}}{M^{4}}+3\frac{M_{2}}{M^{3}}+\frac{5}{2}\frac{M_{2}E^{2}}{M^{5}}-
\frac{3}{2}\frac{E_{2}E}{M^{4}}-\frac{1}{2}\frac{H_{1}^{2}}{M^{4}},\nn\\
Z^{(es)}_{7}&=&11\frac{S_{1}M_{2}}{M^{5}}-6\frac{S_{3}}{M^{4}}+\frac{8}{5}\frac{S_{1}E^{2}}{M^{4}}+
\frac{8}{3}\frac{S_{1}E^{4}}{M^{6}}-\frac{8}{5}\frac{H_{1}E}{M^{3}}-\frac{4}{3}\frac{H_{1}E^{3}}{M^{5}},\nn\\
Z^{(es)}_{8}&=&\frac{153}{28}\frac{S_{1}^{2}}{M^{4}}+\frac{153}{28}\frac{M_{2}}{M^{3}}+
\frac{39}{8}\frac{M_{2}^{2}}{M^{6}}-\frac{15}{4}\frac{M_{4}}{M^{5}}+\frac{81}{14}\frac{M_{2}E^{2}}{M^{5}}+
\frac{25}{6}\frac{M_{2}E^{4}}{M^{7}}-
\frac{15}{4}\frac{E_{2}E}{M^{4}}-3\frac{E_{2}E^{3}}{M^{6}}+\frac{46}{3}\frac{S_{1}^{2}E^{2}}{M^{6}}-{}\nn\\
& &{}-\frac{27}{28}\frac{H_{1}^{2}}{M^{4}}-\frac{H_{1}^{2}E^{2}}{M^{6}}-\frac{26}{3}\frac{H_{1}S_{1}E}{M^{5}},\nn\\
Z^{(es)}_{9}&=&26\frac{S_{1}^{3}}{M^{6}}+31\frac{S_{1}M_{2}}{M^{5}}-15\frac{S_{3}}{M^{4}}+
\frac{34}{15}\frac{S_{1}E^{2}}{M^{4}}+\frac{8}{3}\frac{S_{1}E^{4}}{M^{6}}+
\frac{S_{1}E^{6}}{M^{8}}+32\frac{M_{2}S_{1}E^{2}}{M^{7}}-13\frac{S_{1}E_{2}E}{M^{6}}-
\frac{34}{15}\frac{H_{1}E}{M^{3}}-{}\nn\\
& &{}-\frac{8}{3}\frac{H_{1}E^{3}}{M^{5}}-2\frac{H_{1}E^{5}}{M^{7}}-3\frac{H_{1}^{2}S_{1}}{M^{6}}-\frac{1}{2}\frac{H_{1}E_{2}}{M^{5}}-
\frac{17}{2}\frac{H_{1}M_{2}E}{M^{6}}-10\frac{S_{3}E^{2}}{M^{6}}+\frac{9}{2}\frac{H_{3}E}{M^{5}},\nn
\eea
%%%%%%%%%%%
\bea
\label{wzprosw2}
Z^{(ms)}_{3}&=&2\frac{S_{1}}{M^{2}},\quad
Z^{(ms)}_{4}=\frac{3}{2}\frac{M_{2}}{M^{3}},\quad
Z^{(ms)}_{5}=\frac{E_{1}H}{M^{3}}+2\frac{S_{1}H^{2}}{M^{4}},\nn\\
Z^{(ms)}_{6}&=&7\frac{S_{1}^{2}}{M^{4}}+3\frac{M_{2}}{M^{3}}+\frac{5}{2}\frac{M_{2}H^{2}}{M^{5}}-
\frac{3}{2}\frac{H_{2}H}{M^{4}}-\frac{1}{2}\frac{E_{1}^{2}}{M^{4}},\nn\\
Z^{(ms)}_{7}&=&11\frac{S_{1}M_{2}}{M^{5}}-6\frac{S_{3}}{M^{4}}+\frac{8}{5}\frac{S_{1}H^{2}}{M^{4}}+
\frac{8}{3}\frac{S_{1}H^{4}}{M^{6}}+\frac{8}{5}\frac{E_{1}H}{M^{3}}+\frac{4}{3}\frac{E_{1}H^{3}}{M^{5}},\nn\\
Z^{(ms)}_{8}&=&\frac{153}{28}\frac{S_{1}^{2}}{M^{4}}+\frac{153}{28}\frac{M_{2}}{M^{3}}+
\frac{39}{8}\frac{M_{2}^{2}}{M^{6}}-\frac{15}{4}\frac{M_{4}}{M^{5}}+\frac{81}{14}\frac{M_{2}H^{2}}{M^{5}}+
\frac{25}{6}\frac{M_{2}H^{4}}{M^{7}}-
\frac{15}{4}\frac{H_{2}H}{M^{4}}-3\frac{H_{2}H^{3}}{M^{6}}+\frac{46}{3}\frac{S_{1}^{2}H^{2}}{M^{6}}-{}\nn\\
& &{}-\frac{27}{28}\frac{E_{1}^{2}}{M^{4}}-\frac{E_{1}^{2}H^{2}}{M^{6}}+\frac{26}{3}\frac{E_{1}S_{1}H}{M^{5}},\nn\\
Z^{(ms)}_{9}&=&26\frac{S_{1}^{3}}{M^{6}}+31\frac{S_{1}M_{2}}{M^{5}}-15\frac{S_{3}}{M^{4}}+
\frac{34}{15}\frac{S_{1}H^{2}}{M^{4}}+\frac{8}{3}\frac{S_{1}H^{4}}{M^{6}}+
4\frac{S_{1}H^{6}}{M^{8}}+32\frac{M_{2}S_{1}H^{2}}{M^{7}}-13\frac{S_{1}H_{2}H}{M^{6}}+
\frac{34}{15}\frac{E_{1}H}{M^{3}}+{}\nn\\
& &{}+\frac{8}{3}\frac{E_{1}H^{3}}{M^{5}}+2\frac{E_{1}H^{5}}{M^{7}}-3\frac{E_{1}^{2}S_{1}}{M^{6}}+\frac{1}{2}\frac{E_{1}H_{2}}{M^{5}}+
\frac{17}{2}\frac{E_{1}M_{2}H}{M^{6}}-10\frac{S_{3}H^{2}}{M^{6}}-\frac{9}{2}\frac{E_{3}H}{M^{5}},\nn
\eea
%%%%%%%%%%%
%%%%%%%%%%%
\bea
\label{deprosm1}
A^{(es)}_{2}&=&\frac{1}{3},\quad
A^{(es)}_{3}=0,\quad
A^{(es)}_{4}=-\left(\frac{1}{2}-\frac{2}{9}\frac{E^{2}}{M^{2}}\right),\quad
A^{(es)}_{5}=\frac{20}{9}\frac{S_{1}}{M^{2}},\nn \\
A^{(es)}_{6}&=&-\frac{27}{8}+\frac{M_{2}}{M^{3}}+\frac{3}{2}\frac{E^{2}}{M^{2}}+\frac{1}{3}\frac{E^{4}}{M^{4}},\quad
A^{(es)}_{7}=\frac{28}{3}\frac{S_{1}}{M^{2}}-\frac{28}{9}\frac{H_{1}E}{M^{3}}+
\frac{56}{27}\frac{S_{1}E^{2}}{M^{4}},\nn\\
A^{(es)}_{8}&=&-\frac{225}{16}+\frac{80}{27}\frac{S_{1}^{2}}{M^{4}}+\frac{70}{9}\frac{M_{2}}{M^{3}}+
\frac{20}{9}\frac{M_{2}E^{2}}{M^{5}}+
\frac{15}{2}\frac{E^{2}}{M^{2}}+\frac{5}{9}\frac{E^{4}}{M^{4}}+
\frac{140}{243}\frac{E^{6}}{M^{6}}-\frac{20}{9}\frac{E_{2}E}{M^{4}},\nn\\
A^{(es)}_{9}&=&\frac{81}{2}\frac{S_{1}}{M^{2}}+6\frac{S_{1}M_{2}}{M^{5}}-6\frac{S_{3}}{M^{4}}+
8\frac{S_{1}E^{2}}{M^{4}}+4\frac{S_{1}E^{4}}{M^{6}}-17\frac{H_{1}E}{M^{3}}-4\frac{H_{1}E^{3}}{M^{5}},\nn\\
A^{(es)}_{10}&=&-\frac{6615}{128}+\frac{115}{18}\frac{S_{1}^{2}}{M^{4}}+\frac{935}{24}\frac{M_{2}}{M^{3}}+
\frac{35}{12}\frac{M_{2}^{2}}{M^{6}}-\frac{35}{12}\frac{M_{4}}{M^{5}}+\frac{525}{16}\frac{E^{2}}{M^{2}}+
\frac{70}{81}\frac{E^{6}}{M^{6}}+\frac{770}{729}\frac{E^{8}}{M^{8}}-\frac{140}{9}\frac{E_{2}E}{M^{4}}-{}\nn\\
& &{}-\frac{140}{27}\frac{E_{2}E^{3}}{M^{6}}+\frac{370}{27}\frac{M_{2}E^{2}}{M^{5}}+
\frac{140}{27}\frac{M_{2}E^{4}}{M^{7}}+\frac{700}{81}\frac{S_{1}^{2}E^{2}}{M^{6}}+
\frac{11}{18}\frac{H_{1}^{2}}{M^{4}}-\frac{259}{27}\frac{H_{1}S_{1}E}{M^{5}},\nn\\
A^{(es)}_{11}&=&165\frac{S_{1}}{M^{2}}+\frac{1408}{243}\frac{S_{1}^{3}}{M^{6}}+
\frac{968}{27}\frac{S_{1}M_{2}}{M^{5}}-\frac{352}{9}\frac{S_{3}}{M^{4}}+\frac{418}{27}\frac{S_{1}E^{2}}{M^{4}}+
\frac{440}{27}\frac{S_{1}E^{4}}{M^{6}}+\frac{6160}{729}\frac{S_{1}E^{6}}{M^{8}}+
\frac{616}{27}\frac{S_{1}M_{2}E^{2}}{M^{7}}-{}\nn\\
& &{}-\frac{352}{27}\frac{S_{1}E_{2}E}{M^{6}}-\frac{88}{9}\frac{S_{3}E^{2}}{M^{6}}-\frac{1903}{27}\frac{H_{1}E}{M^{3}}-
\frac{484}{27}\frac{H_{1}E^{3}}{M^{5}}-\frac{616}{81}\frac{H_{1}E^{5}}{M^{7}}-{}\nn\\
& &{}-\frac{88}{9}\frac{H_{1}M_{2}E}{M^{6}}+\frac{22}{9}\frac{H_{1}E_{2}}{M^{5}}+\frac{22}{3}\frac{H_{3}E}{M^{5}}.\nn
\eea
%%%%%%%%%%%
\bea
\label{deprosm2}
A^{(ms)}_{2}&=&\frac{1}{3},\quad
A^{(ms)}_{3}=0,\quad
A^{(ms)}_{4}=-\left(\frac{1}{2}-\frac{2}{9}\frac{H^{2}}{M^{2}}\right),\quad
A^{(ms)}_{5}=\frac{20}{9}\frac{S_{1}}{M^{2}},\nn \\
A^{(ms)}_{6}&=&-\frac{27}{8}+\frac{M_{2}}{M^{3}}+\frac{3}{2}\frac{H^{2}}{M^{2}}+\frac{1}{3}\frac{H^{4}}{M^{4}},\quad
A^{(ms)}_{7}=\frac{28}{3}\frac{S_{1}}{M^{2}}+\frac{28}{9}\frac{H E_{1}}{M^{3}}+
\frac{56}{27}\frac{S_{1}H^{2}}{M^{4}},\nn\\
A^{(ms)}_{8}&=&-\frac{225}{16}+\frac{80}{27}\frac{S_{1}^{2}}{M^{4}}+\frac{70}{9}\frac{M_{2}}{M^{3}}+
\frac{20}{9}\frac{M_{2}H^{2}}{M^{5}}+
\frac{15}{2}\frac{H^{2}}{M^{2}}+\frac{5}{9}\frac{H^{4}}{M^{4}}+
\frac{140}{243}\frac{H^{6}}{M^{6}}-\frac{20}{9}\frac{H_{2}H}{M^{4}}.\nn\\
A^{(ms)}_{9}&=&\frac{81}{2}\frac{S_{1}}{M^{2}}+6\frac{S_{1}M_{2}}{M^{5}}-
6\frac{S_{3}}{M^{4}}+8\frac{S_{1}H^{2}}{M^{4}}+4\frac{S_{1}H^{4}}{M^{6}}+
17\frac{H E_{1}}{M^{3}}+4\frac{E_{1}H^{3}}{M^{5}},\nn\\
A^{(ms)}_{10}&=&-\frac{6615}{128}+\frac{115}{18}\frac{S_{1}^{2}}{M^{4}}+
\frac{935}{24}\frac{M_{2}}{M^{3}}+\frac{35}{12}\frac{M_{2}^{2}}{M^{6}}-
\frac{35}{12}\frac{M_{4}}{M^{5}}+\frac{525}{16}\frac{H^{2}}{M^{2}}+
\frac{70}{81}\frac{H^{6}}{M^{6}}+\frac{770}{729}\frac{H^{8}}{M^{8}}-\frac{140}{9}\frac{H_{2}H}{M^{4}}-{}\nn\\
& &{}-\frac{140}{27}\frac{H_{2}H^{3}}{M^{6}}+
\frac{370}{27}\frac{M_{2}H^{2}}{M^{5}}+
\frac{140}{27}\frac{M_{2}H^{4}}{M^{7}}+\frac{700}{81}\frac{S_{1}^{2}H^{2}}{M^{6}}+
\frac{11}{18}\frac{E_{1}^{2}}{M^{4}}+\frac{259}{27}\frac{E_{1}S_{1}H}{M^{5}},\nn\\
A^{(ms)}_{11}&=&165\frac{S_{1}}{M^{2}}+\frac{1408}{243}\frac{S_{1}^{3}}{M^{6}}+
\frac{968}{27}\frac{S_{1}M_{2}}{M^{5}}-\frac{352}{9}\frac{S_{3}}{M^{4}}+
\frac{418}{27}\frac{S_{1}H^{2}}{M^{4}}+
\frac{440}{27}\frac{S_{1}H^{4}}{M^{6}}+\frac{6160}{729}\frac{S_{1}H^{6}}{M^{8}}+
\frac{616}{27}\frac{S_{1}M_{2}H^{2}}{M^{7}}-{}\nn\\
& &{}-\frac{352}{27}\frac{S_{1}H_{2}H}{M^{6}}-\frac{88}{9}\frac{S_{3}H^{2}}{M^{6}}+\frac{1903}{27}\frac{E_{1}H}{M^{3}}+
\frac{484}{27}\frac{E_{1}H^{3}}{M^{5}}+\frac{616}{81}\frac{E_{1}H^{5}}{M^{7}}+{}\nn\\
& &{}+\frac{88}{9}\frac{E_{1}M_{2}H}{M^{6}}-\frac{22}{9}\frac{E_{1}H_{2}}{M^{5}}-
\frac{22}{3}\frac{H E_{3}}{M^{5}}.\nn
\eea
%%%%%%%%%%%
\end{widetext}
We note that the higher order terms we have computed are by one order lower than the corresponding higher order
terms of Ryan. This is due to computing power restrictions, since as one goes to higher order terms our coefficients 
become far richer in moments than Ryan's. The fact that we have two new sets of moments (the electromagnetic ones)
allows many more combinations of moments in high order terms. Actually, from a practical point of view
these expansions are far more advanced than what will be used in gravitational wave data analysis
in the near future. On the other hand the expressions above present an important feature: in every new order
term a new moment shows up. This suggests that a very accurate observational estimation of the series 
could in principle reveal any moment. 

A glance at the corresponding terms of the two electromagnetic cases shows that each combination of
moments for the ($es$) case is numerically equal to the corresponding combination
for the ($ms$) case, if the electric and magnetic moments are interchanged.
The sign though is the same for combinations of pure electric, or pure  magnetic moments, but
opposite for combinations of electric and magnetic moments. 

The difference by two in the order of the highest computed order term between
the power series for the $\Omega$'s and $\Delta E$ is due to the second derivatives
that appear in Eq.~(\ref{omegas}).

Finally, in order to express $\Delta N$ also as a power series of $v$,
we need to expand $dE_{\textrm{wave}}/dt$ as power series of $v$. As was explained in Sec.~\ref{sec:2.1}
we cannot work out the perturbative analysis of gravitational wave emission at a generic
spacetime background; we can only obtain accurate expressions for $dE_{\textrm{wave}}/dt$
up to $v^4$ after the leading order. However, we know
the numerical factor of the higher order moment appearing at any higher than $v^4$ order.
These higher order moments come from the power series expansion of $\rho$ itself through Eq.~(\ref{dEdtI}) which
describes the main contribution to energy radiation. All other contributions
depend on lower moments at the same order of $v$. Therefore,
in the following formulae for $\rho$ and $dE_{\textrm{wave}}/dt$, we write the power series coefficients
explicitly up to the fourth order, while instead of giving the explicit form of all higher
order coefficients, we give only the higher moment term that occurs at each order of the power expansion.
More specifically, in order to compute the power expansion of $dE_{\textrm{wave}}/dt$ we add up all three
power series contributions of Eqs.~(\ref{dEdtI},\ref{dEdtJ},\ref{dEdtPN}). Thus, we yield
\be
\label{powserrho}
\rho=M v^{-2} \left( 1 + \sum\limits_{n=2}^{\infty} \rho_{n} v^{n} \right),
\ee
and
\be
\label{powserdEdt}
\frac{dE_{\textrm{wave}}}{dt}=\frac{32}{5} \left( \frac{\mu}{M} \right)^2 v^{10}
\left( 1 + \sum\limits_{n=2}^{\infty} W_{n} v^{n} \right),
\ee
where the $\rho_n$ and $W_n$ coefficients for the two electromagnetic cases are respectively
\bea
\label{rhocoefes}
%\rho^{(es)}_{1}  &=&0\nn\\
\rho^{(es)}_{2}  &=&-1-\frac{E^{2}}{M^{2}}\nn\\
\rho^{(es)}_{3}  &=&-\frac{2}{3}\frac{S_{1}}{M^{2}}\nn\\
\rho^{(es)}_{4}  &=&-\frac{1}{2}-\frac{1}{2}\frac{M_{2}}{M^{3}}+\frac{1}{2}\frac{E^{2}}{M^{2}}-\frac{2}{9}\frac{E^{4}}{M^{4}}\nn\\
\rho^{(es)}_{4k+1}&=&-(-1)^{k}\frac{2}{3}\frac{(2k-1)!!}{(2k-2)!!}
\frac{H_{2k-1}E}{M^{2k+1}}  +\textrm{LOM}\nn\\
\rho^{(es)}_{4k+2}&=&-(-1)^{k}\frac{(2k+2)}{3}\frac{(2k-1)!!}{(2k)!!}
\frac{E_{2k}E}{M^{2k+2}}    +\textrm{LOM}\nn\\
\rho^{(es)}_{4k+3}&=&-(-1)^{k}\frac{2}{3}\frac{(2k+1)!!}{(2k)!!}
\frac{S_{2k+1}}{M^{2k+2}}     +\textrm{LOM}\nn\\
\rho^{(es)}_{4k+4}  &=&-(-1)^{k}\frac{1}{3}\frac{(2k+3)!!}{(2k+2)!!}
\frac{ M_{2k+2} }{ M^{2k+3} } +\textrm{LOM}
\eea
%%%%%%%%%%
\bea
\label{rhocoefms}
%\rho^{(ms)}_{1}  &=&0\nn\\
\rho^{(ms)}_{2}  &=&-1-\frac{H^{2}}{M^{2}}\nn\\
\rho^{(ms)}_{3}  &=&-\frac{2}{3}\frac{S_{1}}{M^{2}}\nn\\
\rho^{(ms)}_{4}  &=&-\frac{1}{2}-\frac{1}{2}\frac{M_{2}}{M^{3}}+\frac{1}{2}\frac{H^{2}}{M^{2}}-\frac{2}{9}\frac{H^{4}}{M^{4}}\nn\\
\rho^{(ms)}_{4k+1}&=&(-1)^{k}\frac{2}{3}\frac{(2k-1)!!}{(2k-2)!!}
\frac{E_{2k-1}H}{M^{2k+1}}  +\textrm{LOM}\nn\\
\rho^{(ms)}_{4k+2}&=&-(-1)^{k}\frac{(2k+2)}{3}\frac{(2k-1)!!}{(2k)!!}
\frac{H_{2k}H}{M^{2k+2}}    +\textrm{LOM}\nn\\
\rho^{(ms)}_{4k+3}&=&-(-1)^{k}\frac{2}{3}\frac{(2k+1)!!}{(2k)!!}
\frac{S_{2k+1}}{M^{2k+2}}     +\textrm{LOM}\nn\\
\rho^{(ms)}_{4k+4}  &=&-(-1)^{k}\frac{1}{3}\frac{(2k+3)!!}{(2k+2)!!}
\frac{ M_{2k+2} }{ M^{2k+3} } +\textrm{LOM}
\eea
%%%%%%%%%%%
and
%%%%%%%
\bea
\label{dedt}
%W^{(es)}_1&=&0\nn\\
W^{(es)}_2&=&-\frac{1247}{336}-\frac{4}{3}\frac{E^{2}}{M^{2}}\nn \\
W^{(es)}_3&=&4\pi-\frac{11}{4}\frac{S_{1}}{M^{2}}\nn \\
W^{(es)}_4&=&-\frac{44711}{9072}+\frac{1}{16}\frac{S_{1}^{2}}{M^{4}}
-2\frac{M_{2}}{M^{3}}+6\frac{E^{2}}{M^{2}}-\frac{2}{9}\frac{E^{4}}{M^{4}}\nn\\
W^{(es)}_{4k+1}&=&
-(-1)^{k}\frac{8}{3}\frac{(2k-1)!!}{(2k-2)!!}
\frac{H_{2k-1}E}{M^{2k+1}}  +\textrm{LOM}\nn\\
W^{(es)}_{2k+2}&=&
-4(-1)^{k}\frac{(2k+2)}{3}\frac{(2k-1)!!}{(2k)!!}
\frac{E_{2k}E}{M^{2k+2}}    +\textrm{LOM}\nn\\
W^{(es)}_{4k+3}&=&
-(-1)^{k}\frac{8}{3}\frac{(2k+1)!!}{(2k)!!}
\frac{S_{2k+1}}{M^{2k+2}}    +\textrm{LOM}\nn\\
W^{(es)}_{4k+4}&=&
-4(-1)^{k)}\frac{1}{3}\frac{(2k+3)!!}{(2k+2)!!}
\frac{ M_{2k+2} }{ M^{4k+5} } +\textrm{LOM}
\eea
\bea
%W^{(ms)}_1&=&0\nn\\
W^{(ms)}_2&=&-\frac{1247}{336}-\frac{4}{3}\frac{H^{2}}{M^{2}}\nn \\
W^{(ms)}_3&=&4\pi-\frac{11}{4}\frac{S_{1}}{M^{2}}\nn \\
W^{(ms)}_4&=&-\frac{44711}{9072}+\frac{1}{16}\frac{S_{1}^{2}}{M^{4}}
-2\frac{M_{2}}{M^{3}}+6\frac{H^{2}}{M^{2}}-\frac{2}{9}\frac{H^{4}}{M^{4}}\nn\\
W^{(ms)}_{4k+1}&=&
(-1)^{k}\frac{8}{3}\frac{(2k-1)!!}{(2k-2)!!}
\frac{E_{2k-1}H}{M^{2k+1}}  +\textrm{LOM}\nn\\
W^{(ms)}_{4k+2}&=&
-4(-1)^{k}\frac{(2k+2)}{3}\frac{(2k-1)!!}{(2k)!!}
\frac{H_{2k}H}{M^{2k+2}}    +\textrm{LOM}\nn\\
W^{(ms)}_{4k+3}&=&-(-1)^{k}\frac{8}{3}\frac{(2k+1)!!}{(2k)!!}
\frac{S_{2k+1}}{M^{2k+2}}    +\textrm{LOM}\nn\\
W^{(ms)}_{4k+4}&=&
-4(-1)^{k}\frac{1}{3}\frac{(2k+3)!!}{(2k+2)!!}
\frac{ M_{2k+2} }{ M^{2k+3} } +\textrm{LOM}.
\eea
%%%%%%%%%
In the expressions above the indices $k$ run from 1 to infinity, and the term
LOM is an abbreviation for lower order moments that appear at a specific order
in the expansion. We note that in all these coefficients the same feature with respect to the corresponding terms 
in the two electromagnetic cases that was mentioned before arises.

By combining the power series for $\Delta E/\mu$ (Eq.~(\ref{deprosm}))
with the one for $dE_{\textrm{wave}}/dt$ (Eq.~(\ref{powserdEdt})) we obtain
the power series expansion of $\Delta N$:
\be
\label{powserN}
\Delta N=\frac{5}{96 \pi} \left( \frac{M}{\mu} \right) v^{-5}
\left( 1 + \sum\limits_{n=2}^{\infty} N_{n} v^{n} \right),
\ee
where the $N_n$ coefficients for the two electromagnetic cases are given by the following
polynomials of the moments
\bea
\label{Nfinal}
%N^{(es)}_1&=&0\nn\\
N^{(es)}_2&=&\frac{743}{336}+\frac{14}{3}\frac{E^{2}}{M^{2}}\nn \\
N^{(es)}_3&=&-4\pi+\frac{113}{12}\frac{S_{1}}{M^{2}}\nn \\
N^{(es)}_4&=&\frac{3058673}{1016064}-\frac{1}{16}\frac{S_{1}^{2}}{M^{4}}
+5\frac{M_{2}}{M^{3}}+{}\nn\\
& &{}+\frac{12431}{504}\frac{E^{2}}{M^{2}}+\frac{179}{9}\frac{E^{4}}{M^{4}}\nn\\
N^{(es)}_{4k+1}&=&
(-1)^{k}\frac{(16k+20)}{3}\frac{(2k-1)!!}{(2k-2)!!}
\frac{H_{2k-1}E}{M^{2k+1}}  +{}\nn\\
& &{}+\textrm{LOM}\nn\\
N^{(es)}_{4k+2}&=&
(-1)^{k}\frac{(8k+6)(2k+2)}{3}\frac{(2k-1)!!}{(2k)!!}
\frac{E_{2k}E}{M^{2k+2}}    +{}\nn\\
& &{}+\textrm{LOM}\nn\\
N^{(es)}_{4k+3}&=&
(-1)^{k}\frac{(16k+28)}{3}\frac{(2k+1)!!}{(2k)!!}
\frac{S_{2k+1}}{M^{2k+2}}    +\textrm{LOM}\nn\\
N^{(es)}_{4k+4}&=&
(-1)^{k}\frac{(8k+10)}{3}\frac{(2k+3)!!}{(2k+2)!!}
\frac{ M_{2k+2} }{ M^{2k+3} } +{}\nn\\
& &{}+\textrm{LOM}
\eea
%%%%%%%%%
\bea
\label{Nfinal2}
%N^{(ms)}_1&=&0\nn\\
N^{(ms)}_2&=&\frac{743}{336}+\frac{14}{3}\frac{H^{2}}{M^{2}}\nn \\
N^{(ms)}_3&=&-4\pi+\frac{113}{12}\frac{S_{1}}{M^{2}}\nn \\
N^{(ms)}_4&=&\frac{3058673}{1016064}-\frac{1}{16}\frac{S_{1}^{2}}{M^{4}}
+5\frac{M_{2}}{M^{3}}+{}\nn\\
& &{}+\frac{12431}{504}\frac{H^{2}}{M^{2}}+\frac{179}{9}\frac{H^{4}}{M^{4}}\nn\\
N^{(ms)}_{4k+1}&=&
-(-1)^{k}\frac{(16k+20)}{3}\frac{(2k-1)!!}{(2k-2)!!}
\frac{E_{2k-1}H}{M^{2k+1}}  +{}\nn\\
& &+\textrm{LOM}\nn\\
N^{(ms)}_{4k+2}&=&
(-1)^{k}\frac{(8k+6)(2k+2)}{3}\frac{(2k-1)!!}{(2k)!!}
\frac{H_{2k}H}{M^{2k+2}}    +{}\nn\\
& &{}+\textrm{LOM}\nn\\
N^{(ms)}_{4k+3}&=&
(-1)^{k}\frac{(16k+28)}{3}\frac{(2k+1)!!}{(2k)!!}
\frac{S_{2k+1}}{M^{2k+2}}    +\textrm{LOM}\nn\\
N^{(ms)}_{4k+4}&=&
(-1)^{k}\frac{(8k+10)}{3}\frac{(2k+3)!!}{(2k+2)!!}
\frac{ M_{2k+2} }{ M^{2k+3} } +{}\nn\\
& &{}+\textrm{LOM}.
\eea
%%%%%%%%%

As in the rest three measurable quantities, the power expansion of $\Delta N$
is such that in every order term a new moment, which was not present in any
lower order term, occurs. This proves that all moments can in principle be unambiguously
extracted from accurate measurements of $\Delta N$.

%%%%%%%%%%%%%%%%%%%%%%%%%%%%%%%%%
%%%%%%%%%%%%%%%%%%%%%%%%%%%%%%%%%
\section{\label{sec:4}Using the results in gravitational wave analysis}
%%%%%%%%%%%%%%%%%%%%
%%%%%%%%%%%%%%%%%%%%%%%%%%%%%%%%%

Although we have not quantitatively explored the implications of our results on the
estimation of errors in determining the various moments from a gravitational-wave data analysis,
as it has been done by Ryan in \cite{Ryan2},
we could make some general comments. Actually, the only difference of our
results from the ones of \cite{Ryan} is that more moments are showing up at each coefficient in
the power expansions of all observable quantities, and thus Ryan's estimates for each term
apply equally well here.

As is shown in \cite{Ryan2} the first generation of LIGO is not expected to be able to
extract the first two moments ($S_1$ and $M_2$) with high accuracy ($\sim 0.05$ for the former
and $\sim 0.5$ for the latter one), by analyzing the phase of the waves.
If we allow for electromagnetic fields as well,
the corresponding monopole (which classically is expected to be very close to zero)
will be measured with even higher accuracy than the other two mass moments,
since the charge of the source (or the magnetic monopole in case of some exotic body)
is present at even lower order, namely in the $v^2$ term, while the electric dipole,
or the magnetic dipole, that first show up at the $v^5$ term will be measured with
rather disappointing accuracy. On the other hand, analyzing the data of LISA
leads to accuracies almost two orders of magnitude higher than the corresponding for LIGO.
Thus, it seems quite promising that LISA will give us the opportunity to measure the first few
moments, including the electromagnetic dipole moments, quite accurately. 
Also, the fact that in every new order in the power series of $\Delta N$
a new moment appears is significant, since this means that in principle a unique set of moments
arises from an accurate estimation of all power series terms. Actually, there are two possible sets of
moments; one for each electromagnetic case, since we cannot \textit{a priori} exclude one of them.
We can only exclude one of the two sets on physical grounds, if only one of them leads to
a physically reasonable classical object (for example a highly magnetized compact object
is physically preferable to a compact object with a huge electric dipole). If we manage
to measure a few lower moments, we can check if they are interrelated as in a Kerr-Newman
metric \cite{SotiApos}. A positive outcome of such a test will be of support to the black-hole no-hair theorem
in the case that the central object is a black-hole. The case of observational violation of 
the black-hole no-hair theorem could either mean that the central object is not a black-hole,
or that the theorem does not hold. Of course to assume the latter an extra verification that the central object is
indeed a black hole is necessary. Indications that the central massive compact object is not a black hole
would imply the existence of an exotic object (e.g., soliton star, naked singularity, etc.).
If the central body's mass is measured to be within the stellar limits (e.g., a massive neutron star) we
could get valuable information about its electromagnetic field, like its magnetic dipole field.

While the phase of a gravitational wave is the quantity that can be most accurately
measured, since a large number of cycles (a few thousand for LIGO and a few hundred thousand
for LISA in case of binaries with high-ratio of masses) is sweeping up the sensitive part of the detectors,
the two precession frequencies $\Omega_\rho$ and $\Omega_z$, can in principle be measured
if the detectors become more sensitive and templates that describe modulating waves are used \cite{Apos}.
If this ever become possible one could use any of them to test the no-hair theorem. This 
would demand no more than the four lower order terms, since according to this theorem all moments 
depend on only three quantities (mass, angular momentum, and total charge). 
Actually, from measurements of modulating frequencies we could not at first determine
which frequency corresponds to each precession. However, the power expansions of the two
frequencies begin at a different order, and thus we could discern them.  
Unfortunately, every term of these expansions contains more than one first occurring moment.
However, expansions of $\Omega_\rho$ and $\Omega_z$, if used simultaneously, could lead to
the full determination of the moments.

Our analysis demonstrates that it will be possible to determine
all types of multipole moments of the central object, from future gravitational wave measurements.
Thus, apart from  spacetime geometry, we could also determine the central body's
electromagnetic fields. 
Although the data of LISA should be  suitable for extracting such information with high accuracy, 
the assumptions of circular and equatorial orbit are not that
realistic. From this point of view we consider our work as a step towards a more detailed
analysis with not so restrictive assumptions.

%%%%%%%%%%%%%%%%%%%%%%%%%%%%%
\section*{Acknowledgements}
This research was supported in part by Grant No 70/4/4056 of the Special Account for Research Grants of the
University of Athens, and in part by Grant No 70/3/7396 of the ``PYTHAGORAS'' research funding program.


\begin{thebibliography}{20}
\bibitem{explain} There are two possible electromagnetic fields that lead to reflection symmetric
metric. One of them consists of even electric and odd magnetic moments, while the other one
consists of even magnetic and odd electric moments (see Sec.~\ref{sec:2.3}).
\end{thebibliography}
\end{document}